\documentclass[reprint,superscriptaddress,amsmath,amssymb,aps,pra]{revtex4-2}

\usepackage{graphicx}
\usepackage{dcolumn}
\usepackage{bm}
\usepackage{hyperref}

\usepackage{graphicx}
\usepackage{dcolumn}
\usepackage{bm}

\usepackage[utf8]{inputenc}
\usepackage[T1]{fontenc}
\usepackage{mathptmx}
\usepackage{etoolbox}
\usepackage{{braket}}
\usepackage{hyperref}
\usepackage{tikz}
\usetikzlibrary{arrows,chains,decorations.pathreplacing,decorations.markings}
\usetikzlibrary{quantikz}
\usepackage{placeins}

\usepackage{array}

\DeclareMathOperator*{\argmin}{arg\,min}

\begin{document}

\title{Probabilistic tensor optimization of quantum circuits for the max-$k$-cut problem}
\author{G. V. Paradezhenko}
\affiliation{Skolkovo Institute of Science and Technology, Moscow 121205, Russia}
\author{A. A. Pervishko}
\affiliation{Skolkovo Institute of Science and Technology, Moscow 121205, Russia}
\author{D. Yudin}
\affiliation{Skolkovo Institute of Science and Technology, Moscow 121205, Russia}

\date{\today}

\begin{abstract}
We propose a technique for optimizing parameterized circuits in variational quantum algorithms based on the probabilistic tensor sampling optimization. This method allows one to relax random initialization issues or heuristics for generating initial guess of variational parameters, and can be used to avoid local minima. We illustrate our approach on the example of the quantum approximate optimization algorithm (QAOA) applied to the max-$k$-cut problem based on the binary encoding efficient in the number of qubits. We discuss the advantages of our technique for searching optimal variational parameters of QAOA circuits in comparison to classical optimization methods. 
\end{abstract}

\maketitle

\section{Introduction}
Recent years have witnessed a surge of interest in the benefits of quantum computing~\cite{Gyongyosi2019,Gill2022,Nandhini2022,Sood2023}. With a rather limited number of qubits and relatively short decoherence times, which preclude the widespread use of them, the near-term quantum processors can nevertheless be utilized to solve some combinatorial optimization tasks~\cite{Harrigan2021,Ebadi2022,Yarkoni2022,Nguen2023}, supplementing preceding works in
cold atoms and optical lattices~\cite{Schreiber2015,Choi2016,Bordia2016,Smith2016,Yan2017_1,Bordia2017,Yan2017_2}. Meanwhile, these noisy intermediate scale quantum devices~\cite{Preskill2018} have proved being suitable for the ground state model of computation in the spirit of adiabatic quantum computing~\cite{Aharonov2008}, opening an avenue for solving a wide class of NP-hard problems~\cite{Dam2001}. In combination of both ground state and gate model of quantum computing the ideology of variational quantum computation has been developed~\cite{McClean2016,Cerezo2021,Bharti2022}. This model widely employs a hybrid setup of a quantum processor coupled to an outer loop classical optimization routine. 

In practice, a quantum processor is programmed to prepare a family of parameterized quantum states $\ket{\psi(\bm{\theta})}$, so-called ansatz~\cite{PMS14,Farhi2014}. This ansatz is then measured in the Pauli basis to compute the expectation value of a given Hamiltonian, $E_p(\bm{\theta})$, that is to be minimized. If the class of Hamiltonians is restricted to polynomially bounded cardinality a classical computer can be used to identify expectation values efficiently as based on the measurement statistics. In the following, these values are passed onto an iterative classical optimization routine which variationally adjusts the ansatz to obtain an approximate minimum for the expectation value, $\bm{\theta}_\ast=\argmin \, E_p(\bm{\theta})$. Implemented this way variational quantum algorithms have recently shown several advantages such as robustness to quantum errors and low coherence time requirements, which make them ideal for implementations using available multi-qubit circuits~\cite{McClean2016,Babbush2016,Yang2017,Paesani2017,Li2017,Dunjko2018,Preskill2018,Hempel2018,Colless2018,Santagati2018,Babbush2018,Kivlichan2018,Moll2018,LaRose2019,Schuld2019,Huggins2019,Gyongyosi2019,Cross2019,McArdle2019,Lee2019,Yuan2019,Zhu2019,Wang2019,Liu2019,Carolan2020,Uvarov2020,McArdle2020,Schuld2020,Endo2020,Lubasch2020,Kardashin2020,Cerezo2021,Kardashin2021,Biamonte2021,Monroe2021,Alexeev2021,Harrow2021,Skolik2021}. The main advantage of this methodology is in the fact that one does not need to design a deep quantum circuit~\cite{Malley2016,KMT17,McClean2018}, still the optimization subroutine could be potentially plagued by the presence of barren plateau~\cite{McClean2018}. We herein focus on the quantum approximate optimization algorithm (QAOA)~\cite{farhi2014quantum} for the max-$k$-cut problem and propose a completely different strategy for searching the global minimum of the cost function based on the probabilistic tensor sampling optimization technique~\cite{protes23}. 

In QAOA, optimization tasks have to be specified by a cost Hamiltonian $H_{\rm C}$, a mixing Hamiltonian $H_{\rm M}$, and some initial quantum state $\ket{\psi_0}$. The cost Hamiltonian is typically selected to be diagonal in the computational basis, and its ground state encodes the cost function to be maximized. The mixing Hamiltonian for a set of $n$ qubits $H_{\rm M}$ is typically chosen in the form of the transverse field Hamiltonian, $H_{\rm M} = \sum_{i=j}^n X_j$, where $X_j$ is the $x$ Pauli matrix. The initial state $\ket{\psi_0}$ is any quantum state that can be easily prepared, as a rule of thumb $\ket{\psi_0}  =\ket{+}^{\otimes n}$.

The QAOA circuit of depth $p \geq 1$ is defined as 
\begin{equation}\label{QAOA-circuit}
    U(\bm{\theta}) 
    =  \prod_{k=1}^p U_{\rm M}(\beta_k) \, U_{\rm C}(\gamma_k),
\end{equation}
where the unitaries $U_{\rm C}(\gamma) = e^{-i \gamma H_{\rm C}}$ and $U_{\rm M}(\beta) = e^{-i\beta H_{\rm M}}$ are the phase separation and mixing operators, while $\bm{\theta} = (\gamma_1, \beta_1,\ldots,\gamma_p,\beta_p) \in \mathbb{R}^{2p}$ stands for the variational parameters. Note that a quantum processor is used to prepare the state $\ket{\psi(\bm{\theta})}=U(\bm{\theta})\ket{\psi_0}$. The most pressing issue is to efficiently optimize a parameterized circuit in the non-convex, high-dimensional parameter landscape, where the cost function $E_p(\bm{\theta}) = \bra{\psi(\bm{\theta})} H_{\rm C} \ket{\psi(\bm{\theta})}$ exhibits multiple local minima~\cite{SGY23} and could potentially experience vanishing gradients~\cite{McClean2018}. In QAOA, the choice of a proper optimization technique still represents a challenge as even derivative-free optimizers such as the popular COBYLA~\cite{Powell1994} might be plagued by barren plateaus~\cite{ACC21}. Furthermore, the QAOA performance strongly depends on generating a good initial guess for variational parameters. For $p>1$, various heuristic strategies are developed to subdue this limitation, where an initial guess for $(p+1)$-level QAOA is derived based on optimized parameters from $p$-level~\cite{ZWC20}. These strategies allow one to efficiently find quasi-optimal variational parameters in $O(\mathrm{poly}(p))$ time in contrast to the standard randomized initialization that requires $2^{O(p)}$ runs to obtain an equally good solution. However, these heuristics still are not guaranteed to find the global minimum.

\begin{figure}[t!]
    \centering
    \includegraphics[width=\linewidth]{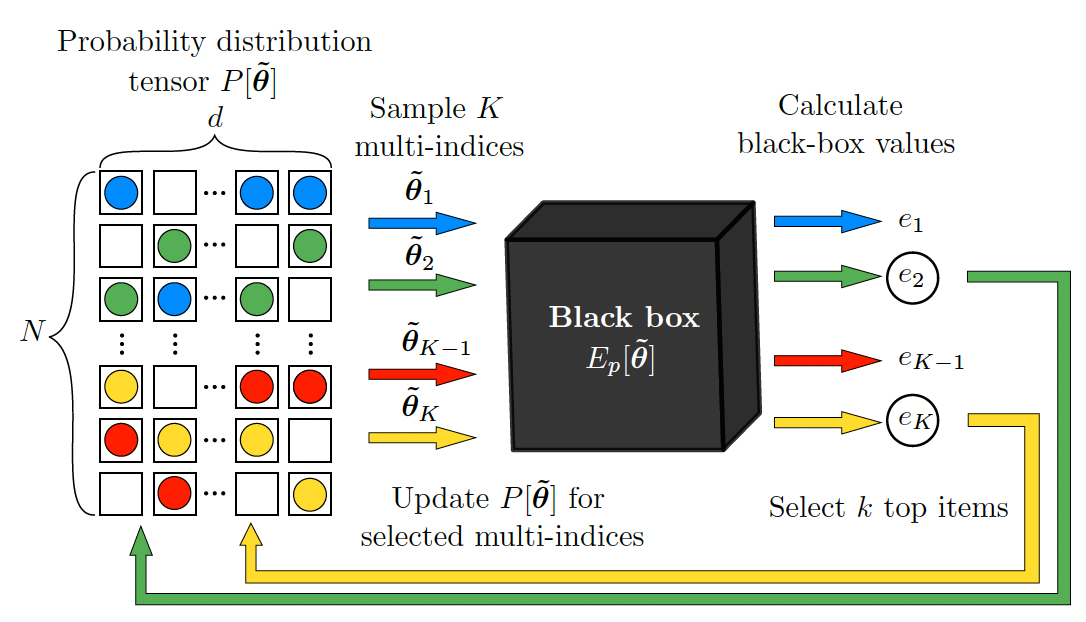}
    \caption{A schematic of the probabilistic tensor sampling method (PROTES)~\cite{protes23}. The probability distribution tensor $P[\bm{\tilde{\theta}}]$ is stored and treated in the tensor-train format.}
    \label{fig:protes}
\end{figure}

\section{Probabilistic tensor sampling}
Since all the parameters in $\bm{\theta}$ are bounded to the interval $[0,2\pi]$, one can introduce a $2p$-dimensional discretization grid for $\bm{\theta}$ in $\Omega = [0,2\pi]^{\times 2p}$ with $N$ nodes per each dimension. We can now consider the function $E_p(\bm{\theta})$ as an implicit $2p$-dimensional tensor $E_p[i_1, i_2,\ldots,i_{2p}]$, where the superscript $(i_1,\,i_2,\ldots,\,i_{2p})$ labels the grid nodes and $1\leq i_k\leq N$,  $k=1,\,2,\ldots,\,2p$. Determining the minimum of the function $E_p(\bm{\theta})$ for $\bm{\theta} \in \Omega$ translates into finding the minimal element of the tensor $E_p[i_1, i_2,\ldots,i_{2p}]$ in the discrete setting. Note that the total number of elements of the tensor $N^{2p}$ scales up exponentially with $p$, making it being hardly evaluated or stored for sufficiently large $p$. However, this issue can be relaxed using the low-rank tensor-train (TT) decomposition. Recall that a $d$-dimensional tensor $P[i_1,i_2,\ldots,i_d] \in \mathbb{R}^{N \times N\times \ldots \times N}$ is said to be stored in the TT-format, if its elements are represented as~\cite{OT10,Ose11}
\begin{eqnarray}
    P[i_1,i_2,\ldots,i_d] & \cong &
    \sum_{r_0=1}^{R_0} \sum_{r_1=1}^{R_1}  \ldots \sum_{r_{d} = 1}^{R_{d}} G_1[r_0,i_1,r_1] \times \nonumber \\ 
    & \times & G_2[r_1,i_2,r_2] \ldots G_d[r_{d-1}, i_d, r_d], \label{TT}
\end{eqnarray}
where $R_0,R_1,\ldots,R_d$ (with convention $R_0 = R_d = 1$) are the TT-ranks, and $G_i \in \mathbb{R}^{R_{i-1} \times N \times R_i}$ are the three-dimensional TT-cores. The TT-format representation~\eqref{TT} of a tensor allows one to store it in a compact low-rank form with the total number of elements being linear in $d$, {\it i.e.}, $d\cdot N \cdot \max_{i=1,\ldots,d} (R_i^2)$. 

The main idea behind the probabilistic tensor sampling optimization (PROTES), as implemented in Ref.~\cite{protes23} and schematically shown in Fig.~\ref{fig:protes}, is to construct a discrete distribution $P[\tilde{\bm{\theta}}]$ of multi-indices $\tilde{\bm{\theta}} = (i_1,i_2,\ldots,i_{2p})$ that samples the minimum $\tilde{\bm{\theta}}_*$ of the tensor $E_p[\tilde{\bm{\theta}}]$ with high probability. The distribution $P[\tilde{\bm{\theta}}]$ is a tensor of the same shape as the objective tensor $E_p[\tilde{\bm{\theta}}]$, but it is treated in the TT-format~\eqref{TT}. In this case, the TT-ranks are assumed to be the same and equal to $R \ll N$, where $R$ is a hyperparameter of the algorithm. To sample the multi-index $\tilde{\bm{\theta}}$ with the probability proportional to the corresponding value $p = P[\tilde{\bm{\theta}}]$ of a tensor represented in the TT-format, the method of sequential calculation of univariate conditional densities with efficient integration in the TT-format is used (for details, see Ref.~\cite{DAF20}). 

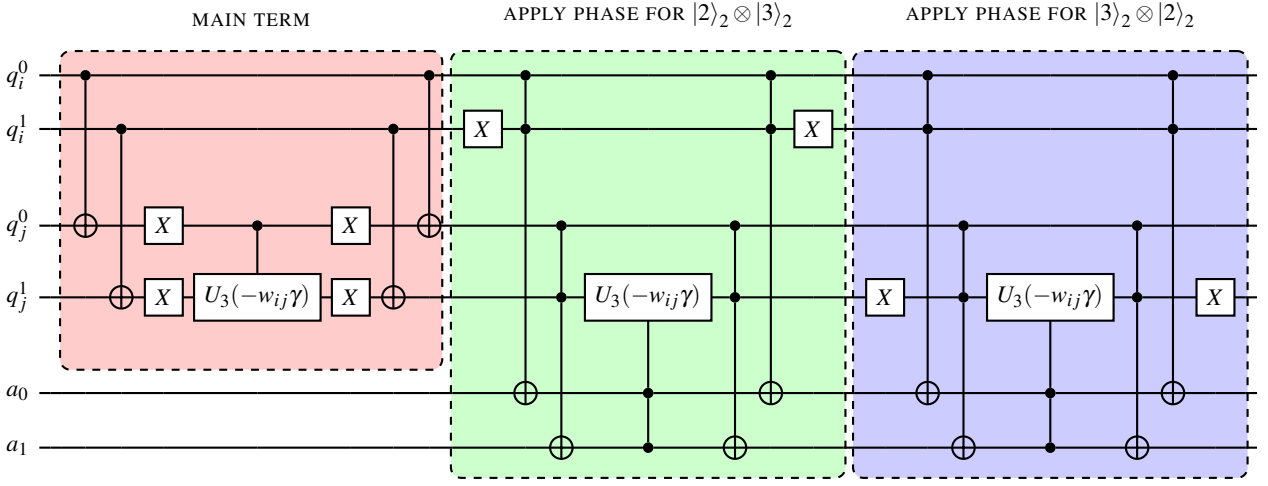
\begin{figure*}[t!]
\centering
\begin{quantikz}[column sep = 0.15cm, row sep = 0.4cm]
\lstick{$q_i^0$} &\qw & \qw 
    \gategroup[6,steps=8,style={dashed,rounded
    corners,fill=red!20, inner
    xsep=-2.5pt},background,label style={label
    position=above,anchor=south,yshift=0.0cm}]{{\sc
    main term}} & \ctrl{3} & \qw & \qw & \qw & \qw & \qw & \ctrl{3} & \qw 
    \gategroup[8,steps=9,style={dashed,rounded
    corners,fill=green!20, inner
    xsep=-2.5pt},background,label style={label
    position=above,anchor=south,yshift=0.0cm}]{{\sc apply phase for $\ket{2}_2 \otimes \ket{3}_2$}} & \qw & \ctrl{1} & \qw & \qw & \qw & \ctrl{1} & \qw & \qw & \qw \gategroup[8,steps=9,style={dashed,rounded
    corners,fill=blue!20, inner
    xsep=-2.5pt},background,label style={label
    position=above,anchor=south,yshift=0.0cm}]{{\sc apply phase for $\ket{3}_2 \otimes \ket{2}_2$}} & \qw & \ctrl{1} & \qw & \qw & \qw & \ctrl{1} & \qw & \qw & \qw &  \\
\lstick{$q_i^1$} &\qw  & \qw & \qw & \ctrl{3} & \qw & \qw & \qw & \ctrl{3} & \qw & \qw & \gate{X} & \ctrl{5} & \qw & \qw & \qw & \ctrl{5} & \gate{X} & \qw & \qw & \qw & \ctrl{5} & \qw & \qw & \qw & \ctrl{5} & \qw & \qw & \qw &   \\
\\
\lstick{$q_j^0$} &\qw  & \qw 
    & \targ{} & \qw & \gate{X} & \ctrl{1} & \gate{X} & \qw & \targ{} & \qw 
    & \qw & \qw & \ctrl{1} & \qw & \ctrl{1} & \qw & \qw & \qw & \qw & \qw & \qw & \ctrl{1} & \qw & \ctrl{1} & \qw & \qw & \qw & \qw \\
\lstick{$q_j^1$} &\qw  & \qw & \qw & \targ{} & \gate{X} & \gate{U_3(-w_{ij}\gamma)} & \gate{X} & \targ{} & \qw & \qw 
    & \qw & \qw & \ctrl{3} & \gate{U_3(-w_{ij}\gamma)} & \ctrl{3} & \qw & \qw & \qw & \qw & \gate{X} & \qw & \ctrl{3} & \gate{U_3(-w_{ij}\gamma)} & \ctrl{3} & \qw & \gate{X} & \qw & \qw &\\
\\
\lstick{$a_0$} & \qw & \qw &  \qw & \qw & \qw & \qw & \qw & \qw & \qw & \qw & \qw & \targ{} & \qw & \ctrl{-2} & \qw & \targ{} & \qw & \qw & \qw & \qw & \targ{} & \qw & \ctrl{-2} & \qw & \targ{} & \qw & \qw & \qw & \\
\lstick{$a_1$}& \qw & \qw &  \qw & \qw & \qw & \qw & \qw & \qw & \qw & \qw & \qw  & \qw & \targ{} & \ctrl{-1} & \targ{} & \qw & \qw & \qw & \qw & \qw & \qw & \targ{} & \ctrl{-1} & \targ{} & \qw & \qw & \qw & \qw &
\end{quantikz}
\caption{Quantum circuit to the phase separation operator~\eqref{U-C-circuit-1} along the edge $(i,j) \in E$ of the graph $G$ implemented as a part of the QAOA parameterized circuit~\eqref{QAOA-circuit} with binary encoding for max-3-cut problem. Qubits $q_i^0$ and $q_i^1$ encode the color assigned to the vertex $i$. Three subsequent terms of the operator~\eqref{U-C-circuit-1} are marked with different colors. For the main term, we apply $CX$-gates on pairs of qubits acting on basis states between the vertices $i$ and $j$. As the result, the state $\ket{q_j}_2$ has only zero entries if and only if qubits $q_i$ and $q_j$ are in the same basis state. Following $X$-gates, we thus apply a controlled $U_3(\phi) \equiv U_3(0,\phi,0)$ gate with $\phi = -w_{ij} \gamma$ only if the original (basis) states $\ket{q_i}_2$ and $\ket{q_j}_2$ are the same. After this, we uncompute by applying $X$ and $CX$-gates in reversed order such that the overall change is that of applying a phase. For the second and third terms in \eqref{U-C-circuit-1}, we apply $CCX$-gates to set two ancillary qubits $a_0$ and $a_1$ to the state $\ket{1}$ if the initial states are $\ket{q_i}_2 = \ket{2}_2$ and $\ket{q_j}_2 = \ket{3}_2$, or vice versa, respectively. If both ancillary qubits are set to $\ket{1}$, a multi-controlled $U_3(\phi)$-gate is applied to change the phase, followed by uncomputation steps. The ancillary qubits are reused for all other pairs $(i,j) \in E$. The multi-controlled $U_3(0,\phi,0)$-gate is implemented in terms of its square root $U_3(0,\phi/2,0)$ and using a polynomial number of $CX$-gates~\cite{YGY07}.}
\label{fig:circuit}
\end{figure*} 

Starting from a random non-negative tensor $P[\tilde{\bm{\theta}}]$ in the TT-format, the following steps are to be implemented until the predefined number of iterations:
\begin{enumerate}
    \item Sample $K$ candidates for $\tilde{\bm{\theta}}_*$ from the current distribution $P[\tilde{\bm{\theta}}]$ and store them as $\Theta_K = \lbrace \tilde{\bm{\theta}}_1, \tilde{\bm{\theta}}_2, \ldots, \tilde{\bm{\theta}}_K \rbrace$;
    
    \item Compute the corresponding values of the cost function: 
    $e_1 = E_p[\tilde{\bm{\theta}}_1]$, $e_2 = E_p[\tilde{\bm{\theta}}_2]$, \ldots, $e_K = E_p[\tilde{\bm{\theta}}_K]$; 

    \item Select $k$ best candidates with indices $S = \lbrace s_1, s_2, \ldots, s_k \rbrace$ from $\Theta_K$ with the minimal values of the cost function, {\it i.e.}, $e_i \leq e_j$ for all $i \in S$ and $j \in \lbrace 1,2, \ldots, K \rbrace / S$;

    \item Update the probability distribution tensor $P[\tilde{\bm{\theta}}]$ to increase the likelihood of selected
    candidates $S$ by making several ($k_{\rm gd}$ in total) gradient ascent steps with the learning rate $\lambda$ for $P[\tilde{\bm{\theta}}]$ using the loss function
    \begin{equation}\label{likehood-loss}
        L(\lbrace \tilde{\bm{\theta}}_{s_1},  \tilde{\bm{\theta}}_{s_2}, \ldots,  \tilde{\bm{\theta}}_{s_k}\rbrace)
        = \sum_{i=1}^k \ln\left( P[\tilde{\bm{\theta}}_{s_i}] \right).
    \end{equation}
\end{enumerate}
Having been completed a sufficient number of iterations, the tensor $P[\tilde{\bm{\theta}}]$ has to have a pronounced peak at the minimal value of the cost function $E_p[\tilde{\bm{\theta}}]$. Note that the objective tensor $E_p[\tilde{\bm{\theta}}]$ is considered as a black-box function, and there is no need to store all its elements during the steps of the algorithm. Instead, we need to calculate only a small part of them (just $K$ elements) on each iteration. Note that the values of $R$, $K$, $k$, $k_{\rm gd}$ and $\lambda$ are the hyperparameters of the algorithm. 

Clearly, the use of PROTES in QAOA does not require an initial point for optimization in contrast to conventional classical optimizers, thus making it possible to avoid any heuristic for generating initial guess when optimizing a QAOA circuit with several layers $p$~\cite{ZWC20}. Moreover, this derivative-free method treats the objective as a black-box function while searching for its global minimum on the bounded domain in a fully probabilistic manner. In practice, we propose to employ a combination of both PROTES and a classical optimizer. Our idea is first to perform a sufficient number of iterations using PROTES to obtain some approximation $\bm{\theta}_*^{(0)}$ to the global minimum in $[0,2\pi]^{\times 2p}$, that is further used as an initial point for a classical optimizer to finalize optimization task and improve its result in the \emph{local} neighbourhood of $\bm{\theta}_*^{(0)}$. To illustrate our methodology in the follow-up analysis we work out the max-3-cut problem. 

\section{Maximum cut}
Consider a weighted undirected graph $G=(V,E)$ uniquely determined by a set of vertices $V = \lbrace 1,2,\ldots,n \rbrace$ and edges $E \subseteq \lbrace (i,j) \, | \, i,j \in V, i \neq j \rbrace $ with non-negative weights $w_{ij} = w_{ji} \geq 0$. In the max-$k$-cut problem, we seek for a partition of $V$ into $k$ subsets such that the sum of the weights of the $k$-cut is maximized,
\begin{equation}\label{cut-mkc}
  C(\bm{x}) = \sum_{(i,j) \in E} w_{ij} \left[ x_i \neq x_j  \right] \to \max_{\bm{x} \in \lbrace 0,1,\ldots,k-1 \rbrace^n},
\end{equation}
where $[\cdot]$ is the Iverson bracket which is equal to 1 if $x_i \neq x_j$, and 0 otherwise. Note in~\eqref{cut-mkc} we introduce a label $x_i \in \lbrace 0,1,\ldots, k-1 \rbrace$ for each vertex $i \in V$ to indicate which partition this vertex belongs to. For randomized approximation algorithms applied for solving the problem \eqref{cut-mkc}, a convenient metric is the approximation ratio,
\begin{equation}\label{approx-ratio}
  \mathbb{E}[C(\bm{x})] \geq \alpha C^*,
\end{equation}
where $C^* = C(\bm{x}^*)$ is the optimal cut for \eqref{cut-mkc}, and $\mathbb{E}$ is the mean value. Clearly, the trivial algorithm that randomly assigns vertices to partitions is characterized by the approximation ratio $\alpha = (1-1/k)$, meaning that the probability for any given edge to possess endpoints in different partitions equals $(1-1/k)$~\cite{FJ97}. For $k=2$, the well-known Goemans-Williamson algorithm~\cite{GW95} based on the relaxation of integer problem \eqref{cut-mkc} into a semidefinite program yields $\alpha = 0.879$. In the following, this algorithm was extended to $k \geq 3$ by Frieze and Jerrum in Ref.~\cite{FJ97}. In particular, for $k=3$ they achieved the approximation ratio $\alpha = 0.833$, which was further improved by De Klerk {\it et al.}~\cite{KPW04} for all $k\geq 3$ with the use of the Lov\'asz $\vartheta$-function. Independently, Geomans and Williamson elaborated on another algorithm for max-3-cut problem that relies on a complex semidefinite program~\cite{GW04}. For $k=3$, they achieved the best-known approximation ratio $\alpha = 0.836$. 

In case of max-2-cut problem, the QAOA was numerically shown to achieve a better approximation ratio in comparison to the classical Geomans-Williamson algorithm~\cite{Cro18}. Note that the approximation ratio~\eqref{approx-ratio} can be viewed as $\alpha = E_p(\bm{\theta}_*)/C^*$. Whereas max-$k$-cut with $k\geq3$~\eqref{cut-mkc} requires a more delicate encoding scheme that is typically chosen out of three available options. Firstly, one can, in principle, use the direct encoding in terms of qudits, representing a $k$-level quantum system (for qudits-based QAOA, see, {\it e.g.}, Refs.~\cite{WUR22,DSL23}). Secondly, one can implement the one-hot encoding, where $k$ bits are used for each vertex $i \in V$, namely a single bit that is equal to 1 encodes a vertex color. This approach requires $nk$ qubits in total and, in addition, the introduction of constraints to guarantee a feasible solution. The latter is done by introducing a penalty term in the phase separation Hamiltonian or using the $XY$-mixer Hamiltonian together with the consistent initial state (for details, see, {\it e.g.}, Refs.~\cite{HWG19,WRD20}). The last but not least, recently developed binary encoding~\cite{FKA21}, where the color $x_i$ of each vertex $i \in V$ is encoded as $\ket{x_i}_L$ with $L = \lceil \log_2 k \rceil$. This approach provides an exponential improvement in terms of the number of qubits, namely are $n\lceil \log_2 k \rceil$ in total. Moreover, it does not require feasibility constraints to be defined into the circuit of the mixer operator. The resulting cost function manifold as specified by the binary encoding might be easier to handle for a classical optimizer due to fewer local minima~\cite{FKA21}. In the following, we focus on the binary encoding approach in the framework of QAOA and apply it for solving the max-3-cut problem.  

We use two qubits per vertex $i \in V$ to identify its color $x_i = \lbrace 0,1,2 \rbrace$ using the set of two-bit strings $\lbrace 00, 01, 10, 11 \rbrace$, where two of them, $10$ and $11$, represent the same color $x_i=2$. For a graph of $n$ vertices, this implementation requires 2$n$ qubits for encoding colors and two ancillary qubits, {\it i.e.}, $2n+2$ in total. The cost Hamiltonian for the binary encoding is given as a sum of local terms, 
\begin{equation}\label{QAOA-binary-cost}
  H_{\rm C} = \sum_{(i,j) \in E} w_{ij}\,H_{ij},
\end{equation}
where $H_{ij}$ is the diagonal matrix modeling the interaction between vertices $i$ and $j$,
\begin{equation}\label{H-ij}
  H_{ij} = \mathrm{diag}(\mathrm{vec} (D)),
  \quad
  D = \begin{pmatrix}
  1 & -1 & -1 & -1\\
  -1 & 1 & -1 & -1\\
  -1 & -1 & 1 & 1\\
  -1 & -1 & 1 & 1
  \end{pmatrix}.
\end{equation}
Here, $\mathrm{vec}()$ is a linear transformation which converts a matrix into a column vector by stacking the columns on top of each other, and $\mathrm{diag}(v)$ is a matrix with the entries of the vector $v$ along its diagonal. The details on the construction of the matrix $D=D^T$ can be found in Ref.~\cite{FKA21}, but it can be represented as a sum of two terms, 
\begin{equation}\label{D-matrix}
    D = (2I - J) + 2\sum_{c,d=2, \, c\neq d}^{3} \Gamma^{\,c,d},
\end{equation}
where $I$ is the identity matrix, $J$ is the matrix of ones and $\Gamma^{\,c,d} \in \mathbb{R}^{4\times 4}$ is a matrix with 1 in the position specified by $c,d$ and zeros otherwise (recall that numeration of elements starts from zero). Clearly, for the mixing operator
\begin{equation}\label{U-M-circuit}
    U_{\rm M}(\beta) = e^{-i \beta H_{\rm M}}
    = e^{-i \beta \sum_{j=1}^{2n} X_j}
    = \prod_{j=1}^{2n} e^{-i \beta X_j}.
\end{equation}
Each multiplier in Eq.~\eqref{U-M-circuit} can be implemented using a single $x$-rotation gate $R_X(\theta) = e^{-i\theta X/2}$. For the phase separation operator, since the diagonal matrices $H_{ij}$ commute, we arrive at
\begin{equation}\label{U-C-circuit}
    U_{\rm C}(\gamma) =  e^{-i \gamma H_{\rm C}}
    = e^{-i \gamma \sum_{(i,j)\in E} w_{ij} H_{ij}}
    = \prod_{(i,j) \in E} e^{-i \gamma \, w_{ij} H_{ij}}. 
\end{equation} 
Taking into account \eqref{H-ij} and \eqref{D-matrix}, we can further decompose the terms of the product in Eq.~\eqref{U-C-circuit} as 
\begin{equation}\label{U-C-circuit-1}
   e^{-i \gamma w_{ij} H_{ij}} 
    = e^{-i \gamma w_{ij} \mathrm{diag}(\mathrm{vec} (2I - J))} 
    \prod_{\substack{c,d = 2 \\ c\neq d}}^3 e^{-2i \gamma w_{ij}\, \mathrm{diag}(\mathrm{vec}(\Gamma^{\,c,d}))},
\end{equation} 
where we used the fact that the involved matrices are diagonal, thus all terms commute. The quantum circuit that implements the phase separation operator~\eqref{U-C-circuit-1} is shown in Fig.~\ref{fig:circuit}. Note that this circuit is constructed as a sequence of three individual blocks for each term in Eq.~\eqref{U-C-circuit-1} depicted by different colors in Fig.~\ref{fig:circuit}. 

The QAOA calculations are implemented using the Qiskit program suite~\cite{qiskit}. The resulting cost function $E_p(\bm{\theta})$ is minimized using PROTES~\cite{protes23}, provided that $R=5$, $K=20$, $k=200$, $k_{\rm gd}=5$ and $\lambda=0.05$, the number of discretization nodes per each dimension $N=100$, and the number of allowed requests to the cost function $m=1000$, followed by classical optimization with COBYLA~\cite{Powell1994} following Ref.~\cite{FKA21} with the total number of function evaluations restricted to $10^6$. 

\begin{figure}[b!]
    \centering
    \includegraphics[width=1.0\linewidth]{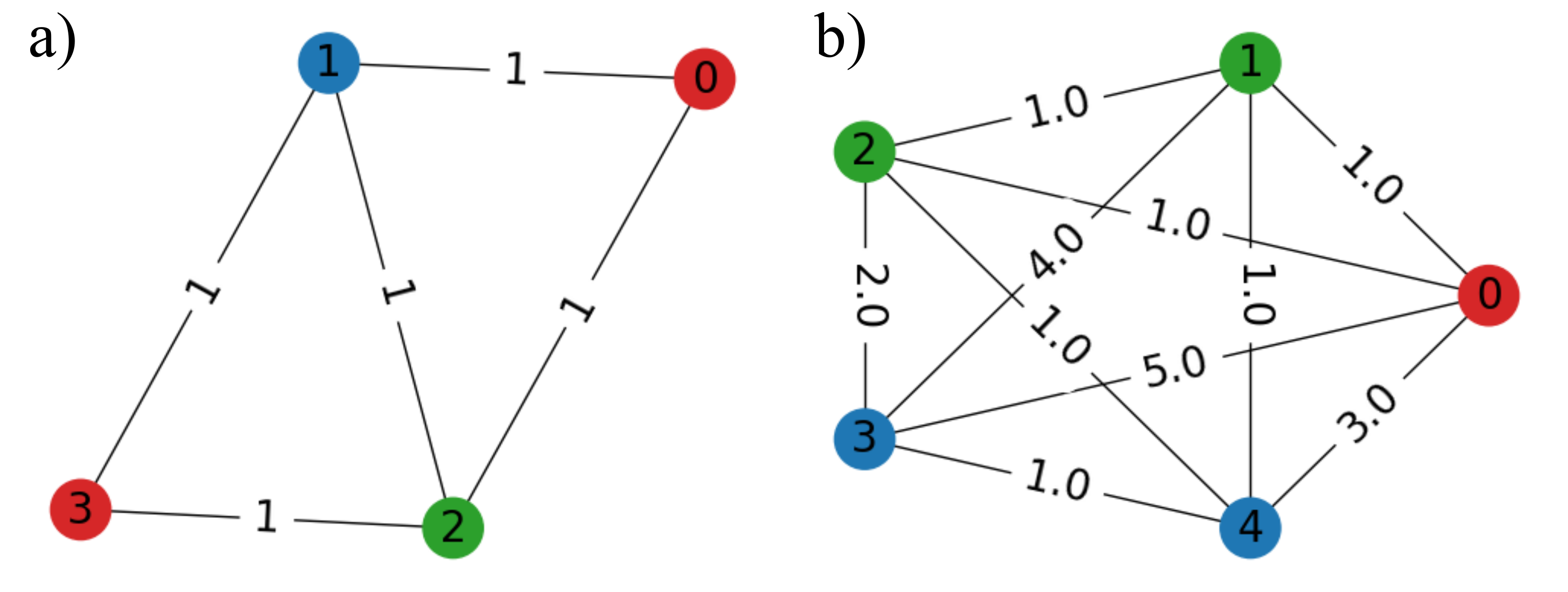}
    \caption{Max-3-cut optimal configurations for a) the complete unweighted graph of $n=4$ vertices, where a single edge is removed;
    b) the complete weighted graph of $n=5$ vertices.}
    \label{fig:graphs}
\end{figure}

\begin{figure}[t!]
    \centering
    \includegraphics[width=0.9\linewidth]{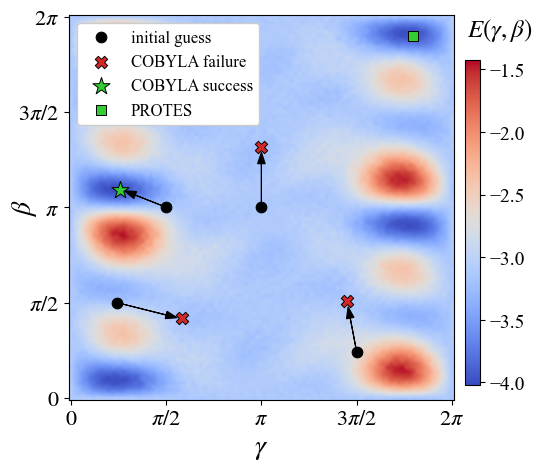}
    \caption{The cost function landscape in case of QAOA with $p=1$ for the unweighted graph of four vertices depicted in Fig.~\ref{fig:graphs}a. The pairs of points connected by arrows show the initial guesses and optimization results as calculated with COBYLA. The success of COBYLA strongly depends on an initial guess. On the contrary, PROTES does not require an initial guess and steadily provides the optimal solution based on the black-box optimization in a fully probabilistic manner.}
    \label{fig:landscape}
\end{figure}

To highlight the advantages of using PROTES, consider a complete unweighted graph of four vertices, where a single edge is removed (see Fig.~\ref{fig:graphs}a). The max-3-cut problem for this graph is simulated using QAOA with a single layer ($p=1$) making it possible to examine the energy landscape of the QAOA cost function versus two variational parameters, $\gamma$ and $\beta$ (see Fig.~\ref{fig:landscape}). As one can see, the QAOA cost function is symmetric with four optimal points in $[0,2\pi] \times [0,2\pi]$. As convenient in the QAOA simulations, the initial guess for optimizing a parameterized circuit by a classical optimizer is generated randomly. However, even for a circuit with a single layer, it strongly affects numerical results as delivered by COBYLA. In other words, if a starting point is chosen far from the solution, COBYLA fails to converge to the minimum. Despite the fact that this method is derivative-free, it still exploits a \emph{local} search direction for minimizing the cost function. Obviously, if it fails to find the optimal solution for $p=1$ the heuristic strategies~\cite{ZWC20} can provide non-optimal solutions for $p>1$, because the initial guess for $(p+1)$-level QAOA is supposed to be based on non-optimal variational parameters calculated for $p$-level. In contrast, PROTES does not depend on the initial guess and searches for the global minimum as a black-box optimizer, {\it i.e.}, it does not employ any local properties of the cost function. As a result, it steadily converges to the minimum in a fully probabilistic manner. 

\section{Numerical results}
\begin{figure}[t!]
    \centering
    \includegraphics[width=0.9\linewidth]{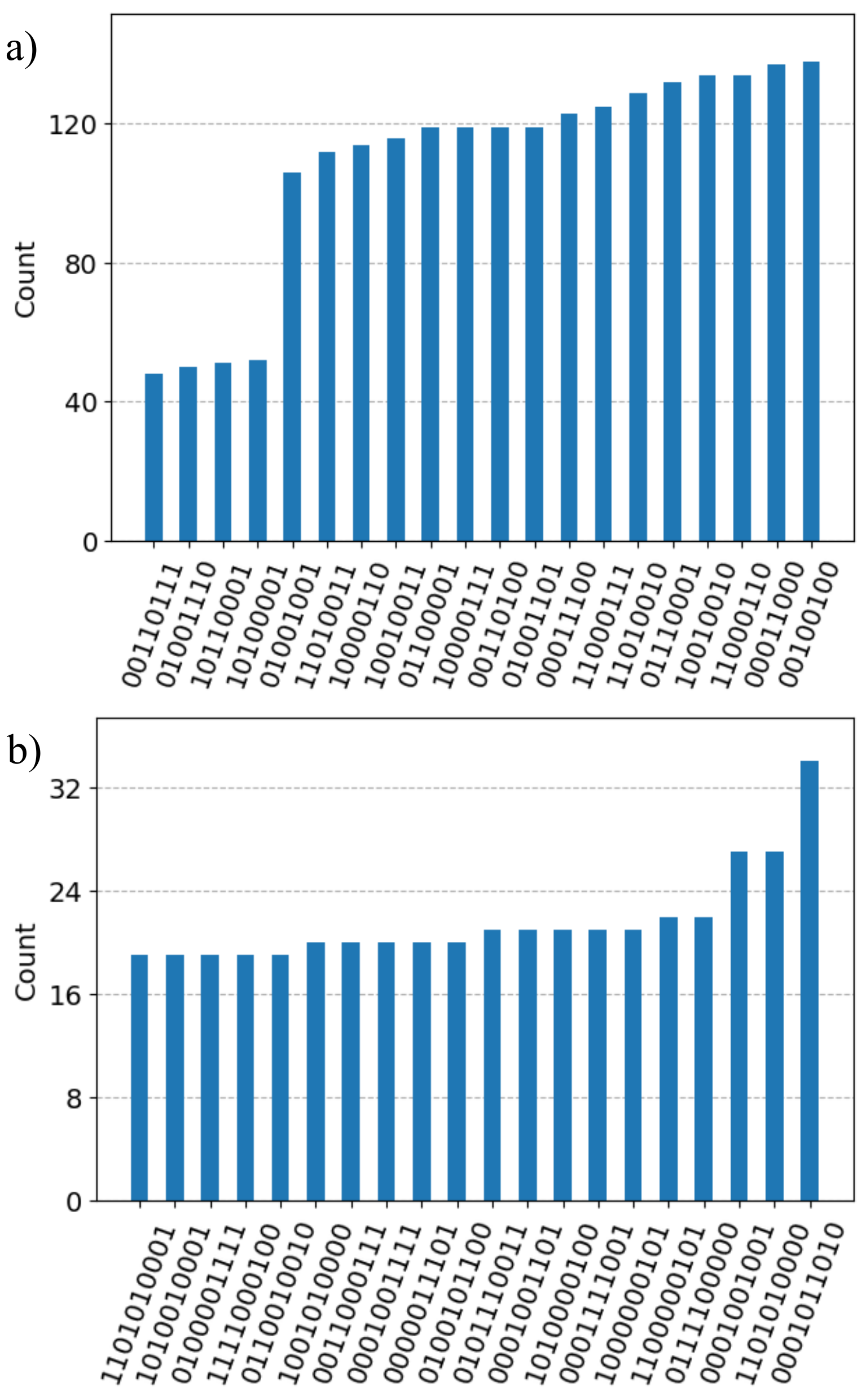}
    \caption{Histograms for counts of bit strings generated by the optimized QAOA circuits with $p=4$ layers for max-3-cut problem with binary encoding. The results are shown for a) the unweighted graph of $n=4$ vertices illustrated in Fig.~\ref{fig:graphs}a; b) the weighted graph of $n=5$ vetrices visualized in Fig.~\ref{fig:graphs}b. The QAOA variational states are optimized based on the consecutive use of PROTES and COBYLA. Note that 4096 shots are utilized for measuring the output, while the histograms show only the first twenty counts of the most frequent measured bit strings. Recall that 10 and 11 encode the same color.}
    \label{fig:hist}
\end{figure}

To proceed, let us discuss the numerical results of the proposed optimization strategy on the example of two graphs shown in Fig.~\ref{fig:graphs}. We simulate max-3-cut problem for both graphs using QAOA with $p=4$ layers. For measuring output bit strings of the QAOA circuit, we use 4096 shots. The achieved approximation ratios are summarized in Table.~\ref{tab:maxcut}, while the statistics of bit strings generated by the optimized QAOA circuits is presented in Fig.~\ref{fig:hist}. For $G_4$, the optimal cut $C^*$ is equal to 5 (see Fig.~\ref{fig:graphs}a). The approximation ratio $\alpha$ achieved with the aid of PROTES is $\alpha_{\rm P} = 0.84$. This result can be further improved up to $\alpha_{\rm C} = 0.87$ by a subsequent use of the COBYLA optimizer. As one can clearly notice, this value is slightly higher as compared to the approximation guarantee~\eqref{approx-ratio} of the classical Goemans-Williamson algorithm, $\alpha_{\rm GW} = 0.84$. As for the measured bit strings, the first sixteen most frequent outputs generated by the optimized QAOA circuit are optimal and almost equiprobable configurations (see Fig.~\ref{fig:hist}a), whereas for the rest (non-optimal) the measurement probability drops expectedly. For $G_5$, the optimal cut $C^*$ is equal to 18 (see Fig.~\ref{fig:graphs}b). In this case, PROTES achieves the approximation ratio $\alpha_{\rm P} = 0.78$, while COBYLA improves this value up to $\alpha_{\rm C} = 0.89$. And, once again, QAOA noticeably outperforms the classical algorithm with $\alpha_{\rm GW} = 0.84$. Similar results for approximation ratios are obtained for different graphs in Ref.~\cite{FKA21}. It is worth mentioning that $G_5$ is characterized by about 20 almost equiprobable configurations as visualized in Fig.~\ref{fig:hist}b. 

\begin{table}[ht!]
\centering
\begin{tabular}{c|c|c|c|c}
graph & $C^*$ & $\alpha_{\rm GW}$ & $\alpha_{\rm P}$ & $\alpha_{\rm C}$ \\ \hline\hline
$G_4$ & 5     & 0.84             & 0.84            & 0.87               \\
$G_5$ & 18    & 0.84             & 0.78            & 0.89               \\ \hline
\end{tabular}
\caption{Max-3-cut numerical results: optimal cut $C^*$, approximation guarantee $\alpha_{\rm GW}$ of the classical Goemans-Williamson algorithm, and approximation ratios $\alpha_{\rm P}$ and $\alpha_{\rm C}$ achieved by QAOA with $p=4$ layers using PROTES exclusively and PROTES followed by COBYLA, respectively.}
\label{tab:maxcut}
\end{table}

\section{Conclusion}
To conclude, we proposed a novel optimization approach for parameterized circuits in the framework of variational quantum algorithms. This approach is based on the probabilistic tensor sampling optimization that allows one to effectively identify the global minimum of multi-dimensional tensors. In comparison to classical optimizers, this method does not require an initial guess, making thus its result independent of random initialization or different heuristics for generating good initial point. Moreover, it is intended to search for the global minimum in a bounded domain using the black-box optimization in a completely probabilistic fashion. The efficiency of the developed approach was demonstrated on the example of QAOA for simulating max-3-cut problem with binary encoding for both weighted and unweighted graphs. Remarkably enough, the resultant approximation ratios are higher as compared to those of the classical Goemans-Williamson algorithm. We believe that our results might be useful for future research to overcome optimization issues in quantum circuits such as a large number of local minima or barren plateaus.

\acknowledgements 
We acknowledge the support from the Russian Science Foundation Project 22-11-00074. We also acknowledge the use of the computational resources at the Skoltech supercomputer ``Zhores''~\cite{Zhores} in our numerical simulations. 

\FloatBarrier

\end{document}